\documentclass[12pt]{article}
\usepackage{times}

\usepackage{epsf}
\def\beq{\begin{equation}}
\def\eeq{\end{equation}}
\def\bea{\begin{eqnarray}}
\def\eea{\end{eqnarray}}
\def\ve{\vert}
\def\vel{\left|}
\def\ver{\right|}
\def\nnb{\nonumber}
\def\ga{\left(}
\def\dr{\right)}

\def\rar{\rightarrow}
\def\nnb{\nonumber}
\def\la{\langle}
\def\ra{\rangle}
\def\ba{\begin{array}}
\def\ea{\end{array}}
\def\bea{\begin{eqnarray}}
\def\eea{\end{eqnarray}}
\def\Bgll{$B_{d}\rightarrow \,\gamma\, \ell^+ \ell^-$}

\title{CP violation in the radiative dileptonic B-meson decays}
\author{\vspace{1cm}\\
         {\bf G\"{u}ray Erkol}
         \thanks{Current address: Kernfysisch Versneller Institute, Zernikelaan 25, 9747 AA Groningen, The Netherlands}
\,\,\thanks{E-mail address:
        erkol@kvi.nl} \, \, and \, \,
        {\bf G\"{u}rsevil  Turan}
        \thanks{E-mail address:
        gsevgur@metu.edu.tr}\,\,}
	 \date{}
\begin{document}
\setlength{\baselineskip}{24pt} \maketitle
\setlength{\baselineskip}{7mm}
~~~~~~~~~~~~~~~~~~~~~~~~~~{\it Middle East Technical University, Physics Dept. Inonu Bul.

~~~~~~~~~~~~~~~~~~~~~~~~~~~~~~~~~~~~~~~~~~~~~~~~~~ 06531 Ankara, TURKEY}

\abstract{We investigate the CP violating asymmetry, the forward backward asymmetry and the CP
violating asymmetry in the forward-backward asymmetry for the
radiative dileptonic B-meson decays \Bgll for the $\ell=~e,\,\mu,\,\tau$ channels. It is
observed that these asymmetries are quite sizable and   \Bgll decays seem promising for  investigating 
CP violation.}
 \thispagestyle{empty} \setcounter{page}{1}
\section{Introduction}
The rare B meson decays are one of the most important probes of the 
effective Hamiltonian governing the flavor-changing neutral current transitions 
$b \rar s(d) \ell^+ \ell^-$. Among them, 
the rare  $B_{s,d} \rar \gamma \, \ell^+ \ell^-$  decays  receive a special interest   due to
their relative cleanliness and sensitivity to new physics. They
have been investigated in the framework of the SM for light and
heavy lepton modes in refs.\cite{Eilam1}-\cite{Dincer}. These decays have also been studied in 
models beyond the SM, such as  MSSM \cite{Xiong} and different versions of  the two Higgs doublet models (2HDM)
\cite{Iltan1}-\cite{Erkol2}, and it was reported that the new physics effects can give sizable
contributions to the relevant observables. 

For $b\rightarrow s\ell^+\ell^-$ transition, the matrix element contains the terms that receive
contributions from $t\bar{t}$, $c\bar{c}$ and $u\bar{u}$ loops,
which are proportional to the combination of
$\xi_t=V_{tb}V^*_{ts}$, $\xi_c=V_{cb}V^*_{cs}$ and
$\xi_u=V_{ub}V^*_{us}$, respectively. Smallness of $\xi_u$
in comparison with $\xi_c$ and $\xi_t$, together with the
unitarity of the CKM matrix elements, bring about the consequence
that matrix element for the $b \rightarrow s\ell^+\ell^-$ decay
involves only one independent CKM factor $\xi_t$, so that the
CP violation in this channel is suppressed  in the SM \cite{Aliev00,Du}.
However, for $b \rightarrow d \ell^+\ell^-$ decay, all the CKM
factors $\eta_t=V_{tb}V^*_{td}$, $\eta_c=V_{cb}V^*_{cd}$ and
$\eta_u=V_{ub}V^*_{ud}$ are at the same order in the SM so that
they can induce a CP violating asymmetry between the decay rates of
the reactions $b \rightarrow d \ell^+\ell^-$  and $\bar{b}\rightarrow \bar{d}\ell^+\ell^-$ 
\cite{Kruger2}. So, $b \rightarrow d \ell^+\ell^-$ decay seems to be suitable 
for establishing CP violation in B mesons. On the other hand, it should be 
noted that the detection of the $b \rightarrow d \ell^+\ell^-$ decay
will probably be more difficult in the presence of a much stronger decay 
$b \rightarrow s \ell^+\ell^-$ and this would make the corresponding exclusive
decay channels more preferable in search of CP violation. In this context, the exclusive
$B_{d} \rar \rho \, \ell^+ \ell^-$  and $B_{d} \rar \pi \, \ell^+ \ell^-$ decays 
have been extensively  studied in the SM \cite{Kruger1} and beyond \cite{Aliev4}-\cite{Durmus}.
So, we think that it would be interesting and complementary to consider the remaining exclusive mode  
$B_{d} \rar \gamma \, \ell^+ \ell^-$.

In this paper, we would like to 
study the CP violation in the exclusive \Bgll decay in 
the context of the SM.  \Bgll decay is induced by the pure-leptonic decay $B_{d} \rar \ell^+ \ell^-$,
which is well known to have  helicity suppression for light lepton modes, having 
branching ratios (BR) of the order of $10^{-15}$ for $\ell=e$ and 
$10^{-10}$ for $\ell=\mu$ channels \cite{Eilam1}. However, 
when a photon line is attached to any of the charged lines in $B_{d} \rar \ell^+ \ell^-$ process, 
it  changes into the corresponding radiative ones, \Bgll, so helicity
suppression is overcome and larger branching ratios are expected.
In \cite{Aliev1}( \cite{Geng}), it was found that in the SM, $BR(B_{d}
\rar \ell^+ \ell^-\gamma)=(1.5 (1.5)~,~1.2 (1.8)~,~ - (6.2))\times 10^{-10}$ for $\ell=e, \mu , \tau$,
respectively. Although these BR's are quite low, in models beyond the SM they
can be enhanced by two (one) orders, as shown e.g. in \cite{Choud2}(\cite{Choud3}) for 
$B_{s(d)} \rar \gamma \, \ell^+ \ell^-$ decay, so investigation of this process may also be
interesting from the point of view of the new physics effects.
 
In \Bgll decays, depending on whether the photon is released from the initial quark
or final lepton lines, there exist two different types of
contributions, namely  the so-called "structure dependent"
(SD) and the "internal Bremsstrahlung" (IB) respectively, while
contributions coming from the release of the free photon from any
charged internal line will be suppressed by a factor of
$m^2_b/M^2_W$. The SD contribution is governed by the vector and
axial vector form factors and it is free from the helicity
suppression. Therefore, it could enhance the  decay rates of the
radiative processes $B_{d} \rar \ell^+ \ell^- \gamma$  in
comparison to the decay  rates of the pure leptonic ones $B_{d}
\rar \ell^+ \ell^-$. As for the IB part of the contribution, it is
proportional to the ratio $m_{\ell}/m_{B}$ and therefore it is
still helicity suppressed for the light charged lepton modes while
it is expected to enhance the amplitude considerably for $\ell=\tau$ mode. 
However, we note that IB part of the amplitude does not contribute to 
CP violating asymmetry $A_{CP}$ and the forward-backward asymmetry $A_{FB}$ (see section 2).

We organized the paper as follows: In section \ref{sect1}, first  the effective 
Hamiltonian is presented and the form factors are defined. Then, the basic formulas
of the differential  branching ratio dBR/dx,  $A_{CP}$, 
$A_{FB}$ and CP violating asymmetry in
forward-backward asymmetry $A_{CP}(A_{FB})$ for \Bgll decay are introduced.
Section \ref{sect2} is devoted to the numerical analysis and discussion.

\section{The theoretical framework of \Bgll decays}\label{sect1} The leading
order QCD corrected effective
Hamiltonian which is induced by the corresponding quark level
process $b \rar d \,  \ell^+ \ell^-$, is given by
\cite{Buchalla}-\cite{Misiak}:
\begin{eqnarray}\label{Hamiltonian} {\cal H}_{eff}  =  \frac{4
G_F}{\sqrt{2}} \, V_{tb} V^*_{td}\Bigg\{ \sum_{i=1}^{10}& \, \,
C_i (\mu ) \, O_i(\mu)-\lambda_u
\{C_1(\mu)[O_1^u(\mu)-O_1(\mu)]\nnb\\&+C_2(\mu)[O_2^u(\mu)-O_2(\mu)]\}\Bigg\}
\end{eqnarray}
where 
\bea\label{CKM}
\lambda_u=\frac{V_{ub}V_{ud}^\ast}{V_{tb}V_{td}^\ast}, 
\eea 
using the unitarity of the CKM matrix i.e.
$V_{tb}V_{td}^\ast+V_{ub}V_{ud}^\ast=-V_{cb}V_{cd}^\ast$. The
explicit forms of the operators $O_i$ can be found in
refs. \cite{Buchalla,Wise}. In Eq.(\ref{Hamiltonian}),
$C_i(\mu)$ are the Wilson coefficients calculated at a
renormalization point $\mu$ and their evolution from the higher scale $\mu=m_W$
down to the low-energy scale $\mu=m_b$ is described by the renormalization group
equation. For $C^{eff}_7(\mu)$ this calculation is performed in refs.\cite{Borzumati,Ciu}
in next to leading order. The value of $C_{10}(m_b)$ to the leading logarithmic approximation 
can be found e.g. in
\cite{Buchalla,Misiak}. We here present the expression for
$C_9(\mu)$ which contains the terms responsible for the CP
violation in \Bgll decay. It has a perturbative
part and a part coming from long distance (LD) effects due to conversion of the
real $\bar{c}c$ into lepton pair $\ell^+ \ell^-$:
\begin{eqnarray}
C_9^{eff}(\mu)=C_9^{pert}(\mu)+ Y_{reson}(s)\,\, ,
\label{C9efftot}
\end{eqnarray}
where
\begin{eqnarray}\label{Cpert}
C_9^{pert}(\mu)&=& C_{9}+h(u,s) [ 3 C_1(\mu) + C_2(\mu) + 3
C_3(\mu) + C_4(\mu) + 3 C_5(\mu) + C_6(\mu) \nonumber
\\&+&\lambda_u(3C_1 + C_2) ] -  \frac{1}{2} h(1, s) \left( 4
C_3(\mu) + 4 C_4(\mu)
+ 3 C_5(\mu) + C_6(\mu) \right)\nnb \\
&- &  \frac{1}{2} h(0,  s) \left[ C_3(\mu) + 3 C_4(\mu) +\lambda_u
(6 C_1(\mu)+2C_2(\mu)) \right] \\&+& \frac{2}{9} \left( 3 C_3(\mu)
+ C_4(\mu) + 3 C_5(\mu) + C_6(\mu) \right) \nonumber \,\, ,
\end{eqnarray}
and
\begin{eqnarray}
Y_{reson}(s)&=&-\frac{3}{\alpha^2_{em}}\kappa \sum_{V_i=\psi_i}
\frac{\pi \Gamma(V_i\rightarrow \ell^+
\ell^-)m_{V_i}}{m_B^2 s-m_{V_i}+i m_{V_i}
\Gamma_{V_i}} \nonumber \\
&\times & [ (3 C_1(\mu) + C_2(\mu) + 3 C_3(\mu) + C_4(\mu) + 3
C_5(\mu) + C_6(\mu))\nnb\\ &+&\lambda_u(3C_1(\mu)+C_2(\mu))]\, .
 \label{Yresx}
\end{eqnarray}
In Eq.(\ref{Cpert}), $s=q^2/m_B^2$ where q is the momentum transfer, $u=\frac{m_c}{m_b}$
 and the functions $h(u, s)$ arise from one loop 
contributions of the four-quark operators $O_1-O_6$ and are given by
\begin{eqnarray}
h(u, s) &=& -\frac{8}{9}\ln\frac{m_b}{\mu} - \frac{8}{9}\ln u +
\frac{8}{27} + \frac{4}{9} y \\
& & - \frac{2}{9} (2+y) |1-y|^{1/2} \left\{\begin{array}{ll}
\left( \ln\left| \frac{\sqrt{1-y} + 1}{\sqrt{1-y} - 1}\right| -
i\pi \right), &\mbox{for } y \equiv \frac{4u^2}{ s} < 1 \nonumber \\
2 \arctan \frac{1}{\sqrt{y-1}}, & \mbox{for } y \equiv \frac
{4u^2}{ s} > 1,
\end{array}
\right. \\
h(0,s) &=& \frac{8}{27} -\frac{8}{9} \ln\frac{m_b}{\mu} -
\frac{4}{9} \ln s + \frac{4}{9} i\pi \,\, . \label{hfunc}
\end{eqnarray}
The phenomenological parameter $\kappa$
in Eq. (\ref{Yresx}) is taken as $2.3$ (see e.g. \cite{Kruger2}).  

Neglecting the mass of the $d$ quark, the effective short distance Hamiltonian
for the $b \rightarrow d \ell^+ \ell^-$ decay in Eq.(\ref{Hamiltonian})  leads to the QCD
corrected matrix element:
\begin{eqnarray}\label{genmatrix}
{\cal M} &=&\frac{G_{F}\alpha}{2\sqrt{2}\pi }V_{tb}V_{td}^{\ast }%
\Bigg\{C_{9}^{eff}(m_{b})~\bar{d}\gamma _{\mu }(1-\gamma _{5})b\,\bar{\ell}%
\gamma ^{\mu }\ell +C_{10}(m_{b})~\bar{d}\gamma _{\mu }(1-\gamma _{5})b\,\bar{%
\ell}\gamma ^{\mu }\gamma _{5}\ell  \nonumber \\
&-&2C_{7}^{eff}(m_{b})~\frac{m_{b}}{q^{2}}\bar{d}i\sigma _{\mu \nu
}q^{\nu }(1+\gamma _{5})b\,\bar{\ell}\gamma ^{\mu }\ell
\Bigg\}.\nonumber\\
\end{eqnarray}
The next step is to calculate the matrix element
of the \Bgll decay. It can be written as the sum of the SD
and IB parts
\bea {\cal M} (B_d \rar
\ell^+ \ell^- \gamma ) & = & {\cal M}_{SD}+ {\cal M}_{IB} \,.\label{calM} 
\eea
It is evident from Eq.(\ref{genmatrix}) that the following matrix elements are needed for the
calculations of ${\cal M}_{SD}$  part :
\bea \label{mel1} 
\la \gamma \vel \bar d \gamma_\mu (1 \mp \gamma_5) b \ver B \ra &=&
\frac{e}{m_B^2} \Big\{ \epsilon_{\mu\nu\lambda\sigma}
\varepsilon^{\ast\nu} q^\lambda k^\sigma g(q^2) \pm i\, \Big[
\varepsilon^{\ast}_{\mu} (k q) - (\varepsilon^\ast q) k_\mu \Big]
f(q^2) \Big\}~, \nnb \eea \bea \la \gamma \ve \bar d i \sigma_{\mu
\nu} q^\nu (1\mp \gamma_5) b \ve B \ra &=& \frac{e}{m_B^2}
\Bigg{\{} \epsilon_{\mu \alpha \beta \sigma} \epsilon^{* \alpha}
q^\beta k^\sigma \, g_1(q^2) \mp ~ i \left[ \epsilon_\mu^* (k q) -
(\epsilon^* q ) k_\mu \right] \,
f_1(q^2) \Bigg{\}}~,\label{ff2} 
\eea 
where $\varepsilon_\mu^\ast$ and $k_\mu$ are the four
vector polarization and four momentum of the photon, respectively,
and $p_B$ is the momentum of the $B$ meson. 
The form factors $g$, $f$, $g_1$, and  $f_1$ were calculated in 
the framework of the light-cone QCD sum rules in \cite{Aliev1,Aliev2,Eilam2}.

The matrix element describing the SD part can be obtained 
from Eqs. (\ref{genmatrix})-(\ref{ff2}),
\bea \label{Msd} 
{\cal M}_{SD} &=& \frac{\alpha G_F}{4
\sqrt{2} \, \pi} V_{tb} V_{td}^* \frac{e}{m_B^2} \,\Bigg\{ \bar
\ell \gamma^\mu (1-\gamma_5) \ell \, \Big[ A_1 \epsilon_{\mu \nu
\alpha \beta} \varepsilon^{\ast\nu} q^\alpha k^\beta + i \, A_2
\Big( \varepsilon_\mu^\ast (k q) -
(\varepsilon^\ast q ) k_\mu \Big) \Big] \nnb \\
&+& \bar \ell \gamma^\mu(1+\gamma_5) \ell \, \Big[ B_1
\epsilon_{\mu \nu \alpha \beta} \varepsilon^{\ast\nu} q^\alpha
k^\beta + i \, B_2 \Big( \varepsilon_\mu^\ast (k q) -
(\varepsilon^\ast q ) k_\mu \Big) \Big] \Bigg\} \, , 
\eea 
where 
\bea
A_1 &=& \frac{-2}{q^2} m_b C^{eff}_7 g_1 + (C^{eff}_9-C_{10}) g ~, \nnb \\
A_2 &=& \frac{-2}{q^2} m_b C^{eff}_7 f_1 + (C^{eff}_9-C_{10}) f ~, \nnb \\
B_1 &=& \frac{-2}{q^2} m_b C^{eff}_7 g_1 + (C^{eff}_9+C_{10}) g ~, \\
B_2 &=& \frac{-2}{q^2} m_b C^{eff}_7 f_1 + (C^{eff}_9+C_{10}) f ~, \nnb ~. 
\eea 
For the IB part, using 
\bea
\la 0 \ve \bar d \gamma_\mu \gamma_5 b \ve B \ra &=& -~i f_B
p_{B\mu}~~~,~~~ \la 0 \ve \bar d \sigma_{\mu\nu} (1+\gamma_5) b \ve B\ra ~=~ 0 ~~,
\eea  
and conservation of the vector current, we get
\bea
\label{Mib} {\cal M}_{IB} &=& \frac{\alpha G_F}{4 \sqrt{2} \, \pi}
V_{tb} V_{td}^* e f_B i \,\Bigg\{ F\, \bar \ell  \Bigg(
\frac{{\not\!\varepsilon}^\ast {\not\!p}_B}{2 p_1 k} -
\frac{{\not\!p}_B {\not\!\varepsilon}^\ast}{2 p_2 k} \Bigg)
\gamma_5 \ell \Bigg\}~,
\eea 
with
\bea
F &=& 4 m_{\ell} C_{10}~~.
\eea 

Substituting Eqs.(\ref{Msd}) and (\ref{Mib}) into Eq. (\ref{calM}), squaring it and then
averaging over the initial and summing over the final spins of the leptons and polarization 
of the photon, we find 
the photon energy distribution, after integration over the phase space, which is given by
\bea \label{dGdx} 
\frac{d\Gamma}{dx} & = & \vel
\frac{\alpha G_F}{4 \sqrt{2} \, \pi} V_{tb} V_{td}^* \ver^2 \,
\frac{\alpha}{\ga 2 \, \pi \dr^3}\,\pi\,m_B \, D(x) 
\eea
where 
\bea \label{Dx} 
D(x) = &&  m_B^2 x^3 v \Bigg[\frac{1}{6}(\vel A_1
\ver^2 + \vel A_2 \ver^2+ \vel B_1 \ver^2 + \vel B_2 \ver^2 )(1-r-x)\nnb \\
&&+ r~Re(A_1~B_1^\ast+A_2~B_2^\ast)\Bigg]+ f_B m_\ell x^2 Re([A_1+B_1]F^\ast)~ln \frac{1+v}{1-v}\nnb\\
&&- f_B^2 \Bigg[2v\frac{1-x}{x}+\Bigg(2+\frac{4r}{x}-\frac{2}{x}-x \Bigg) ln \frac{1+v}{1-v} 
\Bigg] \vel F \ver^2 
\eea
where $x=2 E_{\gamma}/m_B$ is the dimensionless photon energy and $v=\sqrt{1-\frac{4 r}{1-x}}$ with
$r=m^2_{\ell}/m^2_{B}$.

We now consider the CP violating asymmetry, $A_{CP}$, between the \Bgll and
$\bar{B_d}\rar \gamma \, \ell^+ \ell^-$ decays, which  is defined as follows:
\bea A_{CP}(x)& = & \frac{\Gamma (B_d \rar \gamma \, \ell^+ \ell^-)
-\Gamma (\bar{B_d}\rar \gamma \, \ell^+ \ell^-) }{\Gamma (B_d\rar
\gamma \, \ell^+ \ell^-) + \Gamma (\bar{B_d}\rar \gamma \, \ell^+
\ell^-)} ~~. \label{ACP1}
\end{eqnarray}
Using this definition we calculate the $A_{CP}$ as: \bea
A_{CP}=\frac{\int H(x)~dx}{\int (D(x)-H(x))~dx}\eea where
\bea
H(x)=&&\frac{-2x^2}{3}~Im~\lambda_u\Bigg[2~Im~\xi_2\Bigg(C^{eff}_7(f~f_1+g~g_1)m_b~v~x\Bigg(
\frac{x-2r-1}{1-x}\Bigg)\nnb\\&&+ 6 C_{10} g f_B m_\ell^2~ln\frac{1+v}{1-v}\Bigg)-(f^2+g^2)
m_B^2 v x (x-2r-1)~Im~\xi_1^\ast\xi_2\Bigg].
\eea 
In calculating this expression, we use the following parametrization: 
\bea
C_9^{eff}\equiv\xi_1+\lambda_u~\xi_2\, .
\eea

We note that in these integrals the Dalitz boundary for the
dimensionless photon energy $x$ is taken as \bea \delta \leq x
\leq 1-\frac{4 m_\ell^2}{ m^2_B}~, \label{KR} \eea since $\vel
{\cal M}_{IB} \ver^2 $ term has infrared singularity due to the
emission of soft photon. In order to obtain a finite result, we
follow the approach described in ref.\cite{Aliev2} and impose a
cut on the photon energy, i.e., we require $E_{\gamma}\geq 25$
MeV, which corresponds to detecting only hard photons
experimentally. This cut requires that $E_{\gamma}\geq \delta \,
m_B /2$ with $\delta =0.01$.

Next, we  consider the forward-backward asymmetry, $A_{FB}$, in \Bgll.
Using the definition of differential $A_{FB}$
\begin{eqnarray}
A_{FB}(x)& = & \frac{ \int^{1}_{0}dz \frac{d \Gamma }{dz} -
\int^{0}_{-1}dz \frac{d \Gamma }{dz}}{\int^{1}_{0}dz
\frac{d \Gamma }{dz}+ \int^{0}_{-1}dz \frac{d \Gamma }{dz}}\, ,
\label{AFB1}
\end{eqnarray}
where $z=\cos \theta$, $\theta$ is the angle between the momentum of the B-meson and that of $\ell^-$
in the c.m. frame of the dileptons $\ell^+\ell^-$, we find
\begin{eqnarray}
A_{FB}=\frac{\int\, dx\,E(x)}{\int\, dx\,D(x)}\, ,
\label{AFB2}
\end{eqnarray}
with
\begin{eqnarray}
E(x) & = & -4 \, v \, x^2 \Bigg( m^2_B \, x \, \sqrt{(x-1) (x-1+4 r)}\mbox{\rm Re}
(A_1 A^*_2-B_1 B^*_2) \nnb \\
& +& 4 f_B m_{\ell} \, v \, \Bigg(\frac{x-1}{x-1+4 r}\Bigg) \ln
\frac{4 r}{x-1} \mbox{\rm Re}((A_2- B_2) F^*) \Bigg ) \, .\nnb \\ &&
\end{eqnarray}

We have also a CP violating asymmetry in $A_{FB}$, $A_{CP}
(A_{FB})$, which is an important measurable 
quantity in extracting precise information about free parameters of the
models used. Since in the limit of CP conservation, one expects $A_{FB}=-\bar{A}_{FB}$ 
\cite{Buchalla2}, where
$A_{FB}$ and $\bar{A}_{FB}$ are the  forward-backward asymmetries in the particle and 
antiparticle channels, respectively, it is defined as
\begin{eqnarray}
A_{CP}(A_{FB})& = & A_{FB} +\bar{A}_{FB} ~~, \label{ACPAFB1}
\end{eqnarray}
Here, $\bar{A}_{FB}$  can be obtained by the replacement,
\bea
C_9^{eff}(\lambda_u)\rightarrow \bar{C}_9^{eff}(\lambda_u \rightarrow \lambda_u^\ast).
\eea
 
\section{Numerical analysis and discussion}\label{sect2}
\begin{figure}[htb]
\vskip 0truein \centering \epsfxsize=3.8in
\leavevmode\epsffile{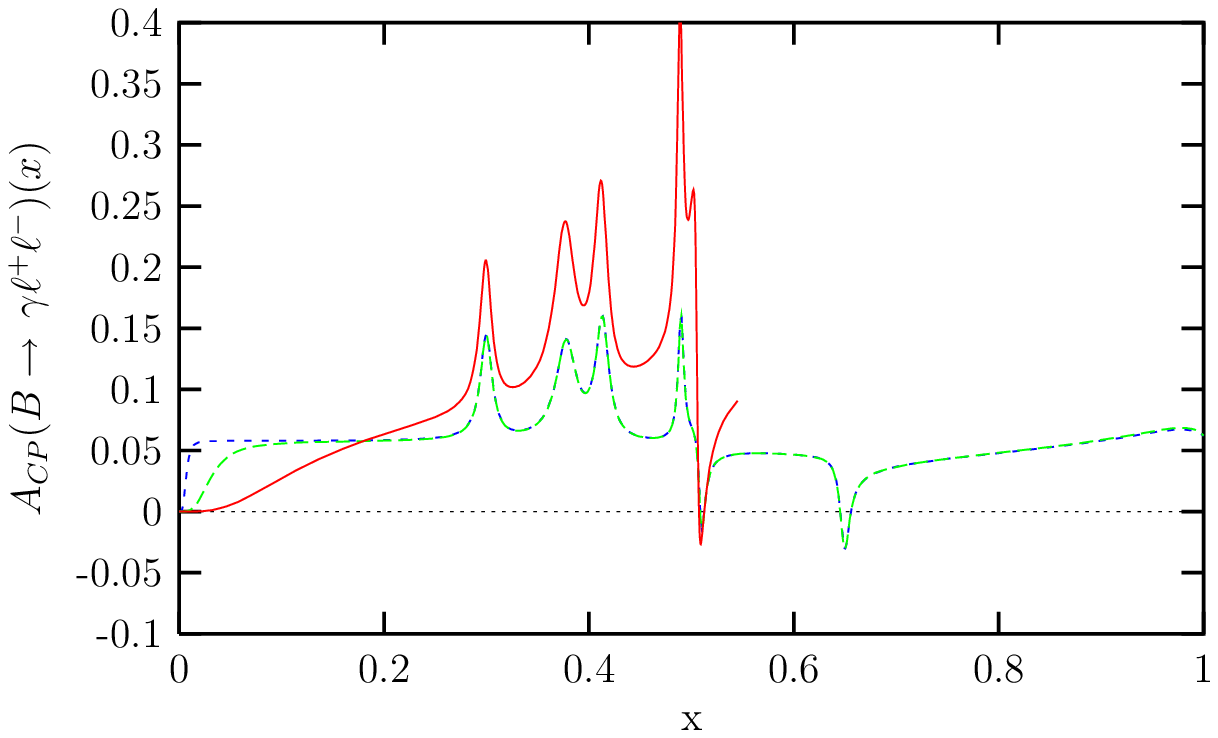} \vskip 0truein
\caption[]{$A_{CP}$ for \Bgll decay for the Wolfenstein parameters
$(\rho,\,\eta)=(-0.07;\,0.34)$. The three distinct lepton modes $\ell=~e,~\mu,~\tau$
are represented by the small dashed, dashed and solid curves,
respectively. } \label{ACP007}
\end{figure}

\begin{figure}[htb]
\vskip 0truein \centering \epsfxsize=3.8in
\leavevmode\epsffile{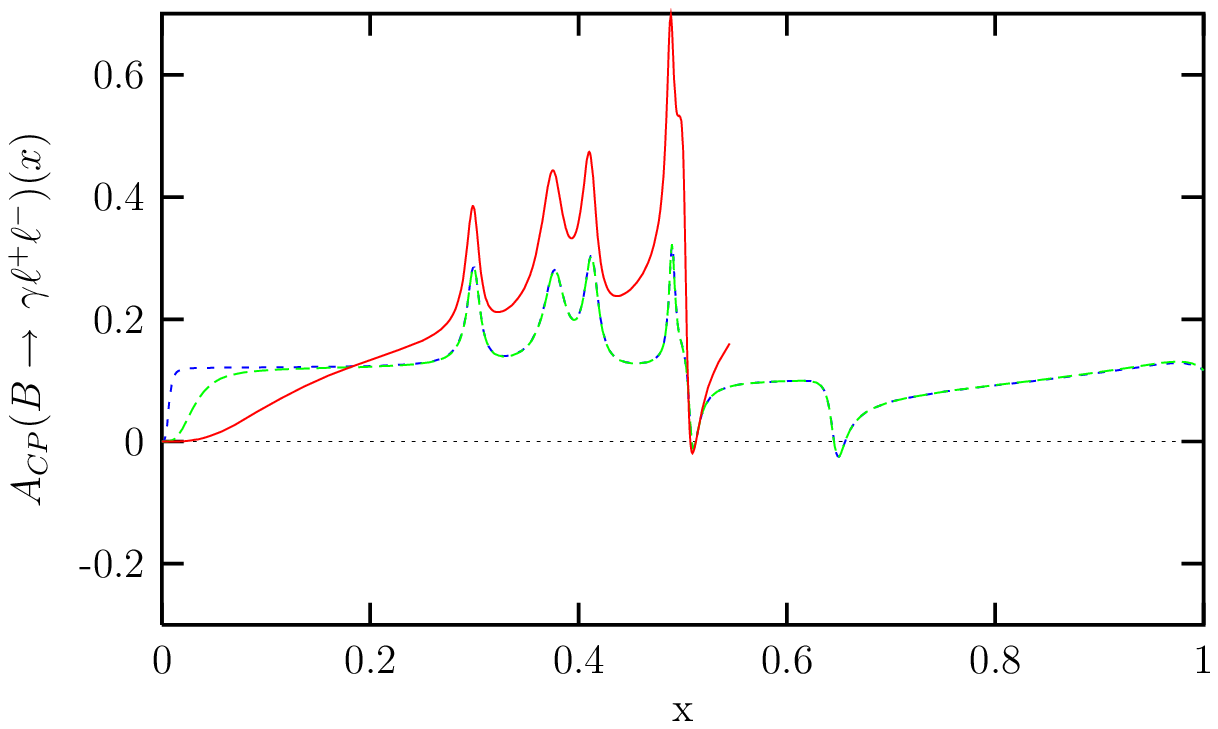} \vskip 0truein \caption[]{The same
as Fig.(\ref{ACP007}) but for the Wolfenstein parameters
$(\rho,\,\eta)=(0.3;\,0.34)$} \label{ACP03}
\end{figure}

\begin{figure}[htb]
\vskip 0truein \centering \epsfxsize=3.8in
\leavevmode\epsffile{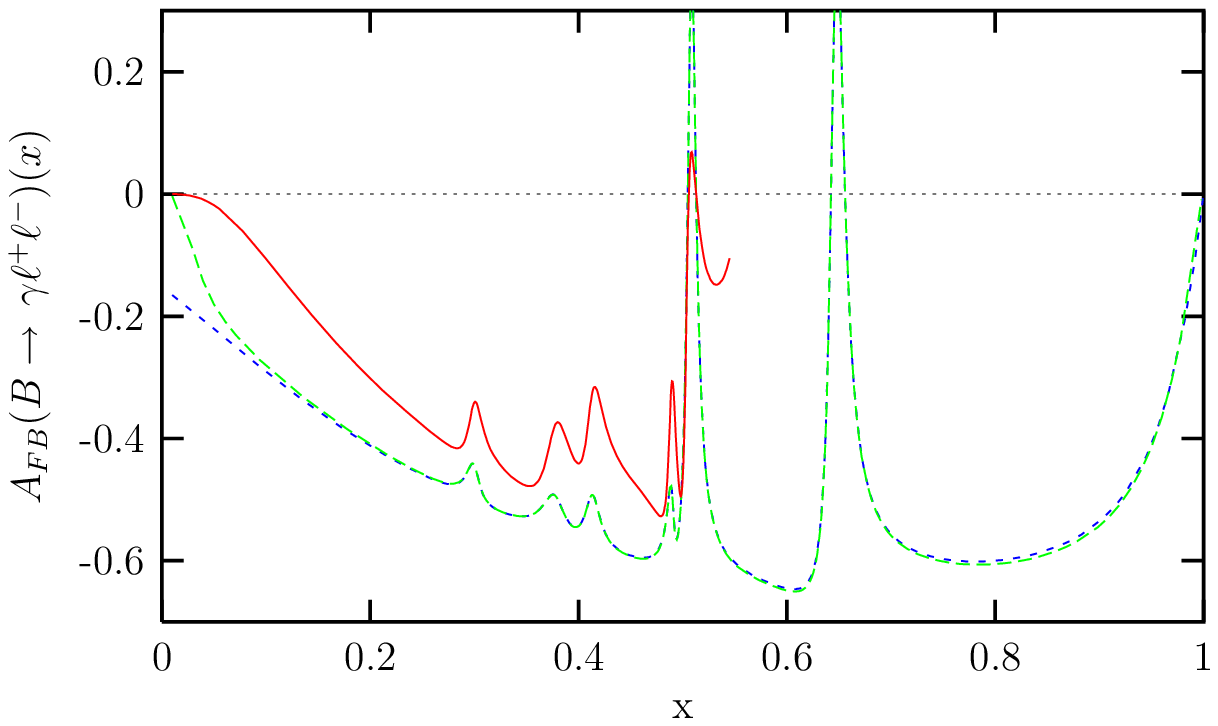} \vskip 0truein \caption[]{$A_{FB}$ for \Bgll decay for the Wolfenstein
parameters $(\rho,\,\eta)=(0.3;\,0.34)$. The three distinct lepton modes $\ell=~e,~\mu,~\tau$
are represented by the small dashed, dashed and solid curves,
respectively.} \label{AFB03}
\end{figure}

\begin{figure}[htb]
\vskip 0truein \centering \epsfxsize=3.8in
\leavevmode\epsffile{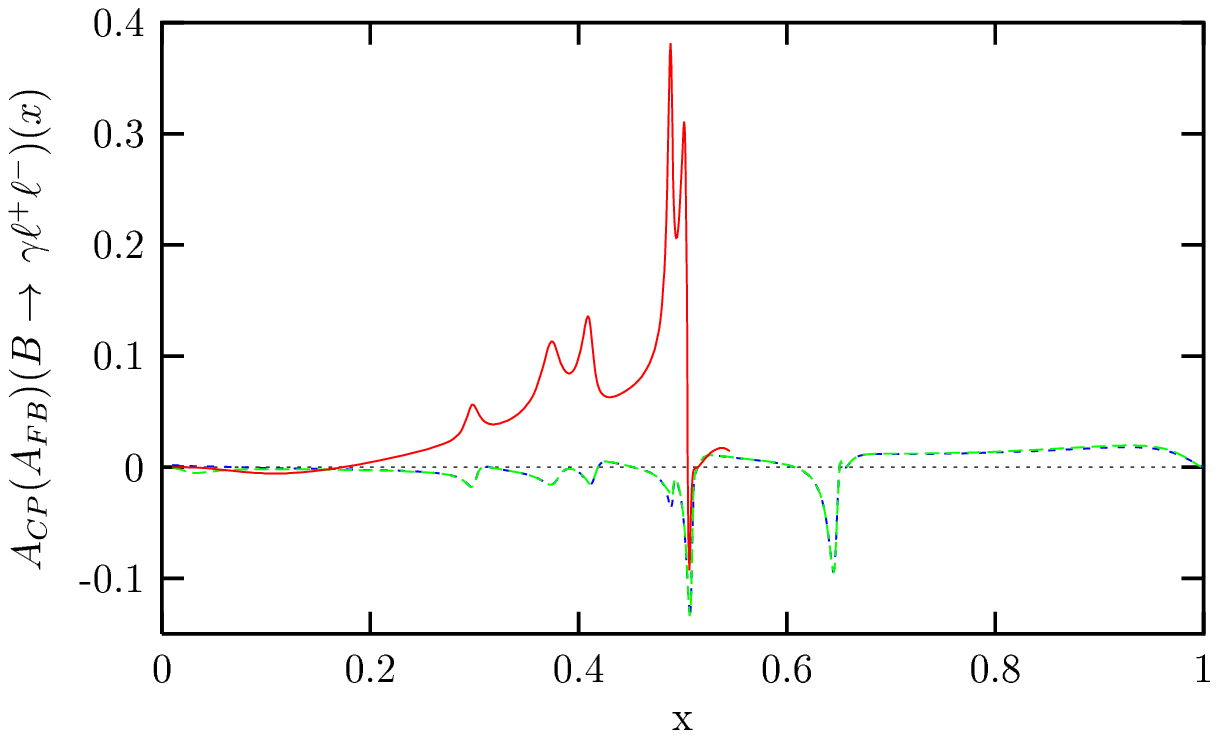} \vskip 0truein
\caption[]{$A_{CP}(A_{FB})$ for \Bgll decay for the Wolfenstein
parameters $(\rho,\,\eta)=(-0.07;\,0.34)$. The three distinct lepton modes $\ell=~e,~\mu,~\tau$
are represented by the small dashed, dashed and solid curves,
respectively.} \label{ACPAFB007}
\end{figure}

\begin{figure}[htb]
\vskip 0truein \centering \epsfxsize=3.8in
\leavevmode\epsffile{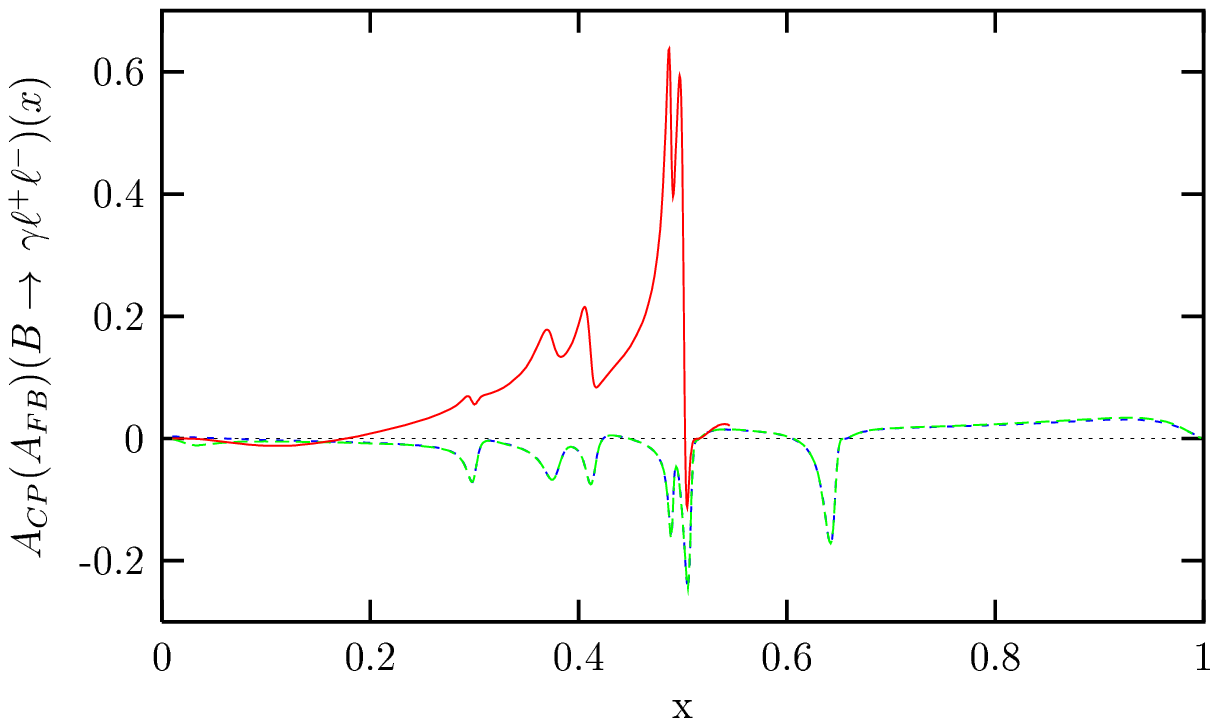} \vskip 0truein \caption[]{The
same as Fig.(\ref{ACP007}) but for the Wolfenstein parameters
$(\rho,\,\eta)=(0.3;\,0.34)$} \label{ACPAFB03}
\end{figure}

In Figs. (\ref{ACP007}-\ref{ACPAFB03}), we present the dependence of the $A_{CP}$, $A_{FB}$ 
and $A_{CP} (A_{FB})$ on the dimensionles photon energy $x$ for the \Bgll $(\ell=e,\,\mu,\,\tau)$ decays
for two different sets of parameters
$(\rho,\,\eta)=(-0.07;\,0.34)$ and $(0.3;\,0.34)$ in the
following Wolfenstein parametrization: 
\bea
\lambda_u=\frac{\rho(1-\rho)-\eta^2-i\eta}{(1-\rho)^2+\eta^2}+O(\lambda^2).
\eea 
We have also evaluated the average values of CP asymmetry $<A_{CP}>$, forward-backward
asymmetry $<A_{FB}>$ and CP asymmetry in the forward-backward asymmetry $<A_{CP} (A_{FB})>$
in \Bgll decay for the above  sets of parameters $(\rho,\,\eta)$, and our results are displayed in Table 1 and
2 with and without including the long distance effects, respectively.

For  the form factors $g,~f,~g_1$ and $f_1$, we have used the values  calculated in 
the framework of light--cone QCD sum rules in refs.  \cite{Aliev1,Aliev2,Eilam2},
which  can be represented in the following dipole forms,
\bea \label{ff} g(q^2) &=&
\frac{g(0)}{\left(1-\frac{q^2}{m_g^2}\right)^2~}, ~~~~~~~~~~
f(q^2) = \frac{f(0) }{\left(1-\frac{q^2}{m_f^2}\right)^2}~,
 \nnb \\
g_1(q^2) &=& \frac{g_1(0)}{\left(1-\frac{q^2}{m^2_{g_1}}\right)^2}~,
~~~~~~~~~
f_1(q^2) = \frac{f_1(0)}{\left(1-\frac{q^2}{m^2_{f_1}}\right)^2}~,
\eea
where
\bea
g(0) & = & 1 \, GeV \, , \,f(0)=0.8 \, GeV \, , \, g_1(0)=3.74 \, GeV^2 \, ,
\,f_1(0)=0.68 \, GeV^2 \,, \nnb \\
m_g & = & 5.6 \, GeV \, , \, m_f=6.5 \, GeV \, , \, m_{g_1}=6.4 \,
GeV \, , \, m_{f_1}=5.5 \, GeV \, . \nnb \eea 
In addition to these form factors, the input parameters and the initial values of the Wilson coefficients we used in our numerical
analysis are as follows:
\begin{eqnarray}
& & m_B =5.28 \, GeV \, , \, m_b =4.8 \, GeV \, , \,m_c =1.4 \,
GeV \, , f_B=0.2 \,GeV \, , \nnb \\
& & m_{\tau} =1.78 \, GeV,\, m_{\mu}=0.105\,GeV,\, \vel V_{tb}V_{td}^\ast\ver=0.01,\,\nnb\\ 
& &C_1=-0.245,\, C_2=1.107,\, C_3=0.011,\, C_4=-0.026,\,C_5=0.007,\,\nnb\\
& & C_{6}=-0.0314,\,C^{eff}_{7}=-0.315,\, C_{9}=4.220, \, C_{10}=-4.619.
\end{eqnarray}

In our numerical analysis, we take into account five possible
resonances for the LD effects coming from the reaction $b \rar d\,
\psi_i \rar d \, \ell^+ \ell^-$, where $i=1,...,5$ and divide the
integration region into two parts for $\ell=\tau$: $\delta \leq x \leq
1-((m_{\psi_2}+0.02)/m_B)^2$ and $1-((m_{\psi_2}-0.02)/m_B)^2 \leq
x \leq 1-(2 m_{\ell}/m_B)^2 $, where $m_{\psi_2}=3.686 $ GeV is
the mass of the second resonance and into three parts for $\ell=e$ and $\mu$: $\delta \leq x \leq
1-((m_{\psi_2}+0.02)/m_B)^2$, $1-((m_{\psi_2}-0.02)/m_B)^2 \leq x \leq 1-((m_{\psi_1}+0.02)/m_B)^2$ and  $1-((m_{\psi_1}-0.02)/m_B)^2 \leq x \leq 1-(2 m_{\ell}/m_B)^2 $, where $m_{\psi_1}=3.097$ GeV is
the mass of the first resonance.

\begin{table}
\begin{center}
\begin{tabular}{|c|ccc|ccc|ccc|}
 \hline\hline
  &&{\scriptsize$<A_{CP}>$}&&&{\scriptsize$<A_{CP}(A_{FB})>$}&&&{\scriptsize$<A_{FB}>$}&\\
 {\scriptsize$(\rho;\,\eta)$}&{\scriptsize$\ell~=~e$}&{\scriptsize$\ell~=~\mu$}&{\scriptsize$\ell~=~\tau$}& {\scriptsize$\ell~=~e$}&{\scriptsize$\ell~=~\mu$} 
 &{\scriptsize$\ell~=~\tau$}&{\scriptsize$\ell~=~e$}&{\scriptsize$\ell~=~\mu$}&{\scriptsize$\ell~=~\tau$}\\
\hline {\scriptsize$(0.3;\,0.34)$}&$0.118$&$0.114$&$0.103$&$-0.004$&$-0.004$&$0.038$&$-0.472$&$-0.460$&$-0.188$\\
\hline
{\scriptsize$(-0.07;\,0.34)$}&$0.061$ & $0.059$&$0.050$&$0.004$&$0.004$&$0.021$&$-0.503$&$-0.490$&$-0.192$\\
  \hline\hline
  \end{tabular}
  \end{center}
  \caption{The average values of
$A_{CP}$, $A_{FB}$ and $A_{CP}(A_{FB})$ in \Bgll for the three distinct
lepton modes including the long distance effects.}\label{tab1}
  \end{table}
\begin{table}
\begin{center}
\begin{tabular}{|c|ccc|ccc|ccc|}
 \hline\hline
  &&{\scriptsize$<A_{CP}>$}&&&{\scriptsize$<A_{CP}(A_{FB})>$}&&&{\scriptsize$<A_{FB}>$}&\\
 {\scriptsize$(\rho;\,\eta)$}&{\scriptsize$\ell~=~e$}&{\scriptsize$\ell~=~\mu$}&{\scriptsize$\ell~=~\tau$}&{ \scriptsize$\ell~=~e$}&{\scriptsize$\ell~=~\mu$} 
 &{\scriptsize$\ell~=~\tau$}&{\scriptsize$\ell~=~e$}&{\scriptsize$\ell~=~\mu$}&{\scriptsize$\ell~=~\tau$}\\
\hline {\scriptsize$(0.3;\,0.34)$}&$0.106$&$0.103$&$0.081$&$0.011$&$0.010$&$0.021$&$-0.520$&$-0.507$&$-0.200$\\
\hline
{\scriptsize $(-0.07;\,0.34)$}&$0.053$ & $0.051$&$0.034$&$0.010$&$0.010$&$0.011$&$-0.529$&$-0.514$&$-0.200$\\
  \hline\hline
  \end{tabular}
  \end{center}
  \caption{The same as Table (\ref{tab1}), but without including the long distance effects.}\label{tab2}
  \end{table}
For reference, we present our SM predictions without long distance effects
\bea
BR(B_d \rar \gamma \ell^+\ell^-) & = & (8.43,8.52,6.22)\times 10^{-10}\, ,
\eea
for $\ell=e,\mu,\tau$. Here, we have used $\tau(B_d)=1.5\times 10^{-12}$ s, $|V_{tb}V^*_{td}|=0.01$,
and $(\rho;\eta)=(0.30;0.34)$. Our result for $\ell=\tau$ is in a good agreement with ref.\cite{Geng}, 
but for $\ell=e$ and $\mu$, they are larger than those of ref.\cite{Aliev1} and \cite{Geng}.

In Fig.(\ref{ACP007}) and Fig.(\ref{ACP03}), we present the
dependence of $A_{CP}$ on the dimensionless photon energy $x$, for
\Bgll decay for the Wolfenstein parameters
$(\rho;\,\eta)=(-0.07;\,0.34)$ and $(\rho;\,\eta)=(0.3;\,0.34)$,
respectively. The three distinct lepton modes $\ell=~e,~\mu,~\tau$
are represented by the small dashed, dashed and solid curves,
respectively. We observe that the $A_{CP}$ for $\ell=~e,~\mu$
cases almost coincide, reaching  up to  $15~\%$  for the intermediate  values of $x$.
The $A_{CP}$ for $\ell=~\tau$ mode exceeds the values of the  other modes and 
reaches $40~\%$ in the high-$x$ region. We also observe from Tables 1 and 2 that 
 including the LD  effects in calculating $<A_{CP}>$ changes the results 
 only $11-15\%$ for $\ell=~e,~\mu$ modes, while $\ell=~\tau$ mode, it is quite sizable,
 $27-47\%$, depending on the sets  of the parameters used for $(\rho;\,\eta)$.

The $x$ dependence of $A_{FB}$ for the \Bgll $(\ell=~e,~\mu,~\tau)$ decays are plotted in 
Fig.(\ref{AFB03}) for $(\rho;\,\eta)=(0.3;\,0.34)$. Since we observe that this dependence 
is almost unchanged for the other set of $(\rho;\,\eta)=(-0.07;\,0.34)$, we do not display it here.
We see that $A_{FB}$ is negative for all values of $x$, except in the resonance regions.
$<A_{FB}>$ amounts to $-50\%$ for $\ell=~e,~\mu$ modes, but it stands smaller for $\ell=~\tau$ mode,
($-20\%$) as expected. The LD effects on $<A_{FB}>$ are about $10\%$, but in reverse
manner, decreasing its magnitude in comparison to the values without LD contributions.

We present the dependence of the $A_{CP}(A_{FB})$ of \Bgll decay
on  $x$ in Fig.(\ref{ACPAFB007}) and
Fig.(\ref{ACPAFB03}) again for two different sets of the
Wolfenstein parameters. As for $A_{CP}$,   $A_{CP}(A_{FB})$ for 
$\ell=~e,$ and $\ell=~\mu$ modes almost coincide. They can have both signs   and  stand smaller 
for all values of $x$ compared to the one of $\ell=~\tau$ mode. For the latter case, 
$A_{CP}(A_{FB})$ is positive for all values of $x$ except in the resonance region and 
it is at the order of magnitude $10^{-1}$. LD effects seem to be quite significant for $<A_{CP}(A_{FB})>$, 
decreasing its value by $60-80\%$ for $\ell=~e,~\mu$ modes, while $\ell=~\tau$ mode,
including LD effects increases $<A_{CP}(A_{FB})>$ by $75-90\%$.

As a conclusion we can say that there is a significant $A_{CP}$
and $A_{CP}(A_{FB})$ for the \Bgll decay, although the branching ratios
predicted for these channels are relatively small because of CKM
suppression. Experimentally, to be able to measure the BR of \Bgll decays, the required number of 
events, N, are $N \sim BR \times (number~of~B-mesons~produced)$. Since
approximately, $6\times 10^{11}$ $B_d$ mesons are expected to be produced per year,
we may hope  that \Bgll  decays could be measured in LHC-B experiment in future.
\newpage

\newpage
\newpage
\newpage
\end{document}